\documentclass{ws-procs975x65}
\usepackage{graphicx}
\usepackage{hyperref}

\def\beq{\begin{equation}}
\def\eeq{\end{equation}}
\begin{document}

\title{Quark-Noave in binaries: \\ Observational signatures and implications to astrophysics}
\author{Rachid Ouyed$^*$, Denis Leahy, Nico Koning}

\address{Department of Physics and Astronomy, University of Calgary,\\
834 Campus Place N.W., Calgary, AB, Canada, T2N 1N4\\
$^*$E-mail: rouyed@ucalgary.ca\\ %, leahy@ucalgary.ca, nakoning@ucalgary.ca\\
%http://www.ucalgary.ca}
}

\author{Jan E. Staff}

\address{Department of Astronomy, University of Florida,\\
P.O. Box 112055, Gainesville, FL, 32611-2055, USA}
%E-mail: jstaff@ufl.edu}

\begin{abstract}
The explosive transition of a massive neutron star to a quark star (the Quark-Nova; QN) releases 
in excess of $\sim 10^{52}$ erg in kinetic energy  which 
 can drastically impact the surrounding environment of the 
QN. A QN is triggered when a neutron star gains
 enough mass to reach the critical value for quark deconfinement to happen in the core.
In binaries,
 a neutron star has access to mass reservoirs (e.g.  accretion from a companion or from  a Common Envelope; CE).
 We explain observed light-curves of  hydrogen-poor superluminous
Supernovae (SLSNe Ia) in the context of a QN occurring in the second CE phase of a massive binary. In particular this model  gives good fits to  light-curves of SLSNe  with double-humped light-curves.  Our model suggests the QN as a mechanism for  CE ejection  and that they be taken into account during binary evolution.
  In  a short period binary with a  white dwarf companion,   the neutron star can quickly grow in mass and experience a QN event. Part of the QN ejecta   collides with the white dwarf; shocking, compressing; and heating it to driving  a thermonuclear runaway    producing a SN Ia impostor (a QN-Ia). 
Unlike ``normal'' Type Ia supernovae where no compact remnant is formed, a QN-Ia produces a  quark star 
undergoing rapid spin-down providing additional power  along with the $^{56}$Ni decay energy.
Type Ia SNe are used as standard candles and contamination of this data by QNe-Ia  can infer an
incorrect cosmology.  
\end{abstract}

\keywords{stars: evolution, supernovae: general, stars: neutron, stars: white dwarfs}

\bodymatter

%%%%%%%%%%%%%%%%% now a standard article style for the most part

\section{Introduction}

 A QN is  the explosive transition of a massive neutron star (NS) to a quark star (QS; the
compact remnant). It  ejects the outermost layers of the NS
as the relativistic QN ejecta with kinetic energy exceeding excess $10^{52}$ erg. The interaction of this ejecta with its surroundings  leads to unique 
phenomena  and has important implications to astrophysics. 
 When occurring in binaries, Quark-Novae (QNe) have the potential to transform our view
 of binary evolution and has serious implications 
 to both high-energy astrophysics and cosmology. 
After a description of the  QN and its energetics  in section 2, we briefly review two cases of QNe in binaries.
The first case is a QN-Ia (section 3) which is  a QN going off in a short period binary consisting of (the exploding) NS and a  white dwarf (WD) which is the mass reservoir. The extremely dense relativistic  QN ejecta impacts (shocks, compresses and heats) the WD and triggers the thermonuclear run-away of a traditional Type Ia.
Along side the type Ia,  the spinning-down  QS provides an additional
power source which ``tampers" with   the energy budget. In the second case,  we  show that a QN occurring in a massive binary can account for the ``exotic" light-cuves of  double-humped hydrogen poor SLSNe (section 4).
We summarize in section 5.

\section{Quark Nova : Overview, energetics and dynamics}

We define $M_{\rm NS, c.}$ as the  critical mass  for a non-rotating NS to undergo quark deconfinement in its  core. 
The presence of enough strange quarks in the
deconfined core of the NS  then triggers the conversion of
hadrons (i.e. matter made of {\it up} and {\it down} quarks) to the conjectured more stable {\it (uds)}  matter  (i.e. matter made of free {\it up}, {\it down} and {\it strange} quarks)\cite{bodmer71,witten84}. In a QN\cite{ouyed2002}, the {\it (ud)-to-(uds)} conversion front  propagates toward the surface of the NS while harnessing  neutrino\cite{keranen}, photon\cite{vogt} and gravitational energy\cite{niebergal,ouyed13b} possibly  yielding
  a detonative regime.   Micro-physics driven hydrodynamical simulations of this
conversion process seem to indicate that a detonation may indeed occur\cite{niebergal} and when coupled with gravitational collapse may lead to a  universal mechanism
 for the ejection of  the NS outermost layers  ($M_{\rm QN}\sim 10^{-3}M_{\odot}$ of  QN ejecta)
with  a universal kinetic energy, $E_{\rm QN, KE}$, of a few times $10^{52}$ erg (i.e. with an associated Lorentz factor exceeding  $\Gamma_{\rm QN}\sim 10$)\cite{keranen,ouyed2009}.  Thus the kinetic  energy released in a QN exceeds that  of a  supernova by at least an order of magnitude.

 The neutron-rich QN ejecta provides a favorable site for \mbox{r-process} nucleosynthesis\cite{jaikumar,kostka1}.  When this ejecta (expanding radially outward from the parent NS)
 collides with the preceding SN ejecta, it re-energizes and re-brightens the SN yielding a superluminous SN\cite{leahy08}. This double-detonation generates  a superluminous double-peaked light-curve if the
 time-delay between the SN and the QN exceeds a few days. We   account  for the luminosity\cite{ouyed12},   the photometric/spectroscopic  signatures\cite{kostka2} as well as introduce nuclear/spallation signatures resulting from the interaction  of the ultra-relativistic  QN ejecta  with the  SN shell and circumstellar material\cite{ouyed2011}.   For shorter time-delays of less than a day, the QN kinetic energy is lost to PdV work but 
  the collision between the  \mbox{r-process-rich} QN ejecta with the SN ejecta yields unique nuclear
 signatures which may explain existing observations\cite{ouyedcas}. The QS shows features
 reminiscent of soft gamma repeaters \cite{ouyed07a,ouyed07b} while the explosion energetics and variability 
 are reminiscent  of   gamma-ray bursts \cite{staff07}. 
 
 When occurring in binaries, the more complex interactions with the companion result in even more interesting
 features.   We review  the 
 key signatures and   main  implications to astrophysics in this paper.

\section{Quark Nova Ia : A QN in a low-mass X-ray binary}

 We first discuss what happens when a NS in a close binary with a WD companion explodes as a QN.
  In this scenario, Roche-Lobe overflow disrupts the WD which produces a Carbon-Oxygen (CO) torus surrounding the 
 NS \cite{ouyed13,ouyed14}.  Alternatively, the NS may fully merge with the WD so that the NS now is in the core of the WD when the QN occurs. The QN will be triggered following sufficient mass accretion.
 % from the disrupted WD. 

 Some of the relativistic QN ejecta will impact  (shock, heat and compress) the disrupted
  WD inducing a runaway nuclear burning of the CO
in an event we termed a QN-Ia since it is   ``Type Ia''-like explosion.  A
  crucial difference  here however is the  QS which provides extra power through 
  magnetic braking spin-down and   consequently a QN-Ia (which spectrum resembles a Type-Ia SN) is powered by a combination of $^{56}$Ni decay and the spin-down luminosity of the QS. This has drastic consequences
  for cosmological models  if QNe-Ia contaminate
 the sample of Type Ia SNe used as distance indicators in cosmology as discussed below.

\subsection{Implication for cosmology}

The spin-down contribution yields a {\it red-shift-dependent Phillips-like relation}
(\cite[Figure 1]{ouyed14} shows the correlation between peak absolute magnitude and light-curve shape)  which means that they can confuse (i.e. are {\it NOT} rejected by the)  light-curve fitters 
used for cosmology (\cite[Figure 4]{ouyed14}).     
 The rate of QNe-Ia may be a significant fraction of the observed  Type Ia SNe and may  be dominant
at higher redshift\cite{ouyed14}.  This is especially egregious given that the QN-Ia light-curve
 varies with redshift.   

To estimate the effect of contamination, we analyzed hundreds of synthetic QNe-Ia light-curves 
using   the SALT2 light-curve fitting software\cite{guy07} to find the difference ($\Delta \mu(z)$) between the actual distance modulus and the fitted distance modulus as a function of redshift, $z$. Most of the simulated QNe-Ia were best fitted\cite{ouyed14} with :
\begin{equation}
\Delta \mu(z) = -1.13655~e^{-0.39035z} + 1.32865\ .
\end{equation}
For $z=0$ there is a strong correlation since $\Delta \mu(z)\simeq 0$, but at $z=1.5$ the correlation is much weaker $\Delta \mu \sim 0.7$.
We conclude that if QNe-Ia represent an important  fraction of the SNe used in the work which  estimates the accelerating expansion of the Universe\cite{perlmutter99,riess98}, this may have drastically altered the statistics and conclusions  of those studies \cite{ouyed14}.  It is thus vital to differentiate between QNe-Ia and standard SNe-Ia.
Applying our correction above to the Union2.1 data\cite{union2.1} we obtain the true distance moduli of the observed SNe-Ia (if they are indeed QNe-Ia) as shown by red crosses in \cite[Figure 6]{ouyed14}.    This  demonstrates how the SNe-Ia distance moduli, when corrected, lay very close to the $\Omega_{M} =1$, $\Omega_{\Lambda} = 0$ curve.

A QN-Ia may have  already  been observed in SN 2014J \cite{ouyed15sn2014j}. 
SN 2014J's  $^{56}\mathrm{Ni}$ mass was estimated to be $\sim0.36~{\rm M_\odot}$ \cite{churazov14}, based on $^{56}\mathrm{Co}$ decay lines.
However, based on the peak luminosity, $\mathrm{M_{Ni}} \sim 0.77~{\rm M_\odot}$ is expected.
 In the QN-Ia model, the discrepancy can be accounted for by the  QS spin-down power.
 Perhaps the best prospect for observationally distinguishing a QN-Ia from a SN-Ia is
 the detection of  the gravitational wave signal  produced during the explosive transition of the NS to a QS. If the QN is asymmetric, it should emit a gravitational wave signal that could be observable by Advanced LIGO\cite{staff12}. This would be followed by another signal from the exploding WD; here the  time delay between the QN and the exploding WD   is the time it takes the QN ejecta to reach the 
disrupted WD plus the burning time of the WD.
Other types of Type Ia SNe (i.e. single or double degenerate scenarios) would  lack such a dual signal.
 Another strong observational signature  of a QN-Ia is high-energy emission (specifically X-ray signatures\cite{ouyed07a,ouyed07b}) from the QS which would be
impossible in standard SNe-Ia channels. The QS, being an aligned rotator, would  be  radio quiet \cite{ouyed06}. These
signature may be observed in the near future in SN 2014J.

%\begin{verbatim}
\begin{figure}[t!]
\centering
\includegraphics[width=3in]{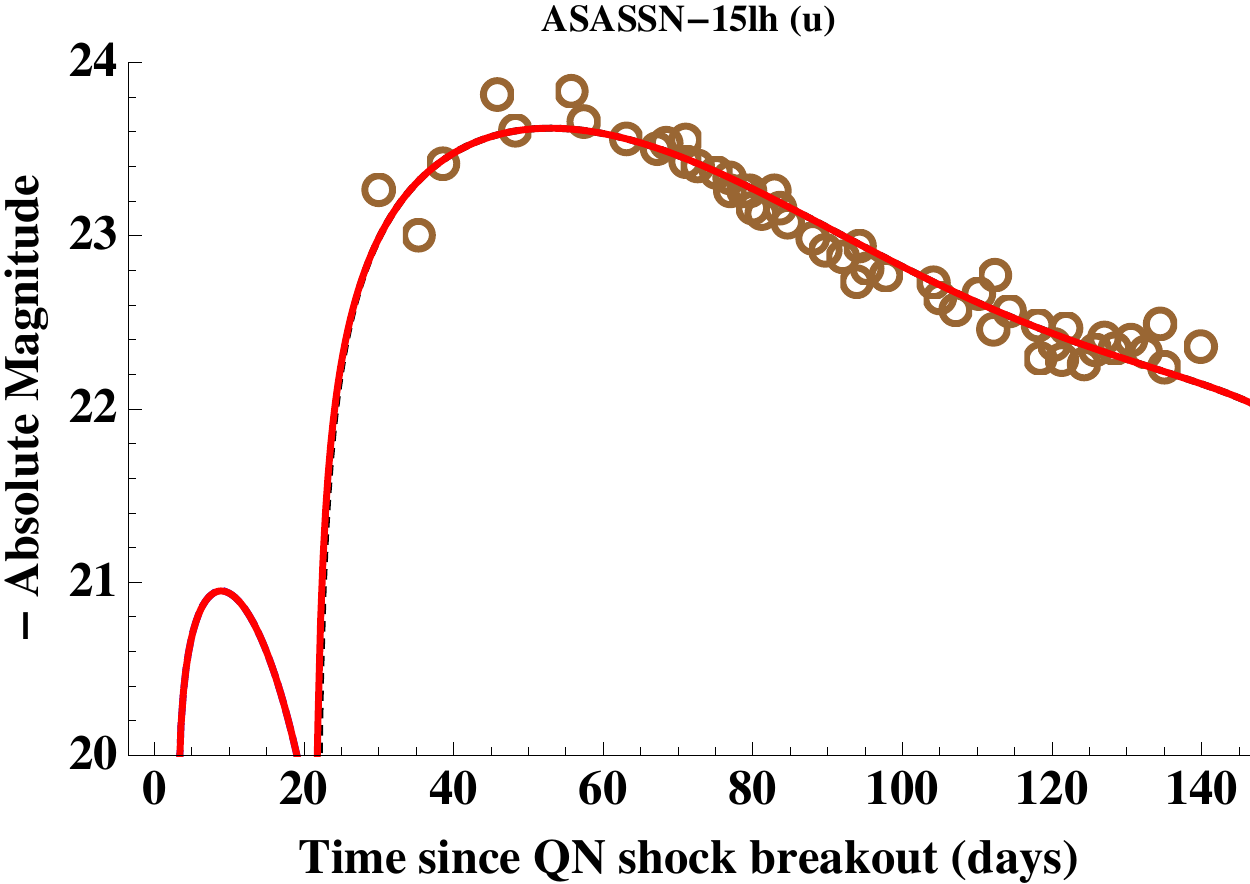}
\caption{A QN occurring inside the second, hydrogen-poor,  CE  of a massive binary: The first hump results from the ejection and energization of the He-rich CE  ($M_{\rm CE}=5M_{\odot}; R_{\rm CE}=3000R_{\odot}$) by the QN. The second  hump is from the BH-accretion 
phase after the QS merges with the CO core.  The circles show u-band observations of
 the  SLSN ASASSN-15lh\cite{dong2016}.}
\label{aba:fig1}
\end{figure}
%\end{verbatim}

\section{Quark Novae in Massive Binaries : a model for SLSNe Ia}

 QNe are likely to occur in binaries experiencing CE phases where the NS
can accrete enough mass to reach $M_{\rm NS, c.}$. In particular, in massive binaries experiencing
two CE phases, the NS would have access to two mass reservoirs\cite{ouyed15,ouyed16}.
 In this picture, the NS would have evolved earlier from its more massive progenitor and is companion to
 the second star during its CE phase.

When the 
 system  enters the CE phase (in which the hydrogen envelope is  ejected), it leaves behind a close binary consisting of the NS and the giant star's He core. At this point the He-core has a mass of a few $M_{\odot}$ and an orbital separation of $\sim 3~{\rm R_\odot}$ (\cite[Figure 1]{ouyed15}).
During this phase, the NS  accretes up to $\sim0.1~{\rm M_\odot}$  and relies on the second CE to grow in mass.
The He core then expands causing a  second He-rich CE phase.
Sufficient mass is accreted onto the NS during this second CE and the NS reaches
$M_{\rm NS, c.}$ and undergoes a QN explosion inside the expanded Hydrogen-poor envelope.
The QN ejecta shocks and unbinds the  CE providing a 
 bright, short-lived hump matching those observed in double-humped  SLSNe\cite{ouyed15,ouyed16}.

Following the QN, the remaining system consists of a QS and the CO core (of mass $M_{\rm CO}$) of the He star.  Orbital decay
lead to a merger a few days to a few weeks following the QN event.
The QS then rapidly accretes from the CO core leading to the  collapse of the QS into a  Black Hole (BH).
The remainder of the CO core  is subsequently accreted by the BH.
The accretion luminosity powers the long lasting main hump of the double-humped SLSN.
 Figure 1 shows our fit to the recently discovered SLSN ASASSN-15lh\cite{dong2016}. We fit it with a CE mass and radius of $5M_{\odot}, 3000R_{\odot}$, respectively. The BH-accretion
parameters are $2\times 10^{46}$ erg s$^{-1}$ for the initial accretion luminosity with an injection
power in time $\propto t^{-1.5}$. The time delay between the QN event and the
onset of BH-accretion  is 20 days.  These parameters are similar to those
 we used to fit a number of hydrogen-poor SLSNe\cite{ouyed16}  (\cite[Table 2]{ouyed16}  and \cite[Figure 2]{ouyed16}). Our model can also fit double-peaked SLSNe showing late-time emission (e.g.
iPTF13ehe and LSQ14bdq) which we modelled as the  collision between the He-rich CE (ejected by the QN) and the hydrogen-rich (i.e. first) CE ejected during the firts CE phase (\cite[Figure 1]{ouyed16}).  The available accretion energy $\eta_{\rm BH}M_{\rm CO}c^2$ ($\eta_{\rm BH}$ is the BH-accretion efficiency) is enough to account for the 
 extreme radiation released during the long-lasting hump in SLSNe.
The QN is key to our model since besides accounting for the first peak, 
it also ejects the second CE at speeds of a  few $10,000$ km s$^{-1}$ which ensures a
very efficient harnessing of the BH-accretion input power by the very
large envelope a few days to a few weeks following the QN event.  The QN deposits its momentum and
energy impulsively in the CE which makes our model fundamentally different from those
involving spin-down power where the energy is deposited gradually.

\section{Summary}

QNe should be  common in binaries   where
 accretion onto the NS from a companion (i.e. the disrupted WD in LMXBs) or during a CE phase 
 (i.e. during  massive binaries evolution)  can drive the NS above the critical
 mass, $M_{\rm NS, c.}$, triggering the QN.  The ability of the QN model in binaries to fit SLSNe in general (see http://www.quarknova.ca/LCGallery.html) and in particular a number of double-humped  SLSNe Ia, suggests that QNe may be an
 important component of  massive binary evolution and may even be responsible for  CE ejection. 
 The QN-Ia  model has two main interesting features:  First, the detonation of the WD in the QN-Ia scenario is explained by standard shock physics governing the interaction of the QN ejecta and the WD.  Secondly, the QN-Ia provides an elegant explanation for the correlation between peak magnitude and light-curve shape through the contribution of spin-down energy to the  light-curve.  Our model  can be tested by further work including  simulations of 
 QNe in binary evolution.
 
 Our model relies on the feasibility of the QN explosion which requires sophisticated simulations of the burning of a NS to a QS which are being pursued. Preliminary simulations with consistent treatment of nuclear and neutrino reactions,  particle diffusion and hydrodynamics show instabilities which  could lead to a detonation\cite{niebergal}.	We also propose  that a ``core-collapse" QN could  result from the collapse of the quark matter core\cite{ouyed13b} which provides another avenue for the explosion. 

\section*{Acknowledgments}

The research of RO, DL and NK is funded by the Natural Sciences and Engineering Research Council of Canada.
J.E.S is funded by the University of Florida Theoretical Astrophysics Fellowship.

\end{document}